\documentclass[10pt,journal,compsoc]{IEEEtran}
\ifCLASSOPTIONcompsoc
  \usepackage[nocompress]{cite}
\else
  \usepackage{cite}
\fi
\ifCLASSINFOpdf
\else
\fi
\usepackage{amsmath,amsfonts}
\usepackage{algorithmic}
\usepackage{algorithm}
\usepackage{array}
\usepackage{textcomp}
\usepackage{stfloats}
\usepackage{url}
\usepackage{verbatim}
\usepackage{graphicx}
\usepackage{cite}
\usepackage{multirow}
\usepackage{pifont}
\usepackage{booktabs}
\usepackage{color}
\usepackage{xcolor}
\usepackage{ulem} 
\usepackage{makecell}
\usepackage{threeparttable}
\usepackage{diagbox}
\usepackage{fancyhdr}

\definecolor{mygreen}{RGB}{112, 173, 71}
\definecolor{myred}{RGB}{192, 0, 0}
\newcommand{\cmark}{\color{mygreen}{\ding{51}}}
\newcommand{\xmark}{\color{myred}{\ding{55}}}

\newcommand{\tool}[1]{{\textsf{\small{#1}}}}

\newcommand{\taitou}[1]{\vspace{0.1cm} \noindent \textbf{{#1.}}}

\usepackage{xspace}
\newcommand{\sysname}  {\textsc{EvilScreen}\xspace}

\begin{document}

\title{EvilScreen Attack: Smart TV Hijacking via Multi-channel Remote Control Mimicry}

\author{Yiwei Zhang, 
        Siqi Ma,
        Tiancheng Chen,
        Juanru Li, 
        Robert H. Deng, 
        Elisa Bertino
        }

\IEEEtitleabstractindextext{%
\begin{abstract}
Modern smart TVs often communicate with their remote controls (including those smart phone simulated ones) using multiple wireless channels (e.g., Infrared, Bluetooth, and Wi-Fi).
However,
    this multi-channel remote control communication introduces a new attack surface. 
An inherent security flaw is that remote controls of most smart TVs are designed to work in a benign environment rather than an adversarial one,
    and thus wireless communications between a smart TV and its remote controls are not strongly 
    protected.
Attackers could leverage such flaw to abuse the remote control communication and compromise smart TV systems. 

In this paper, we propose \sysname, a novel attack that exploits ill-protected remote control communications to access protected resources of a smart TV or even control the screen.
\sysname exploits a multi-channel remote control mimicry vulnerability present in today smart TVs.
Unlike other attacks, which compromise the TV system by exploiting code vulnerabilities or malicious third-party apps,
    \sysname directly reuses commands of different remote controls,
    combines them together to circumvent deployed authentication and isolation policies,
    and finally accesses or controls TV resources remotely.
We evaluated eight mainstream smart TVs and found that they are all vulnerable to \sysname attacks,
    including a \tool{Samsung} product adopting the ISO/IEC security specification.
\end{abstract}

}

\maketitle

\IEEEdisplaynontitleabstractindextext

%
\IEEEpeerreviewmaketitle


\IEEEraisesectionheading{\section{Introduction}}
\IEEEPARstart{S}{mart} TVs present both privacy and security risks. 
Features such as Internet-based media playing and third-party app executing make modern TVs smarter and yet more vulnerable to security attacks and privacy intrusions.
A variety of vulnerabilities have been exploited against smart TVs in recent years~\cite{bachy2015smart, bachy2019smart, ghiglieri2017exploring, kang2017obtain, michele2014watch, mohajeri2019watching, niemietz2015not, sidiropoulos2013smart}.
In general, security threats against smart TVs can be classified into two categories:
    threats from Internet, and threats from programs running on smart TV OSes
    (e.g., Android TV OS~\cite{android_tv}).
In response,
    smart TV manufacturers and TV OS providers have deployed a variety of protection measures. 

\begin{figure}[!t]
  \centering
  \includegraphics[width=0.5\textwidth]{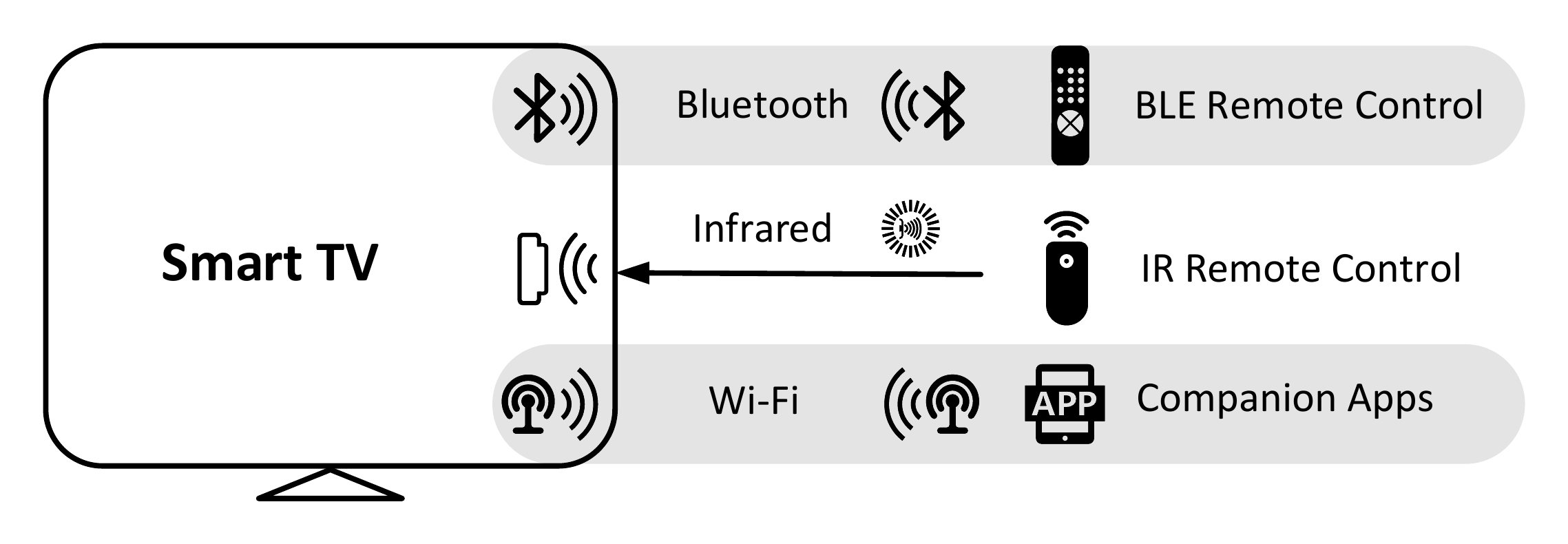}    
  \caption{A common multi-channel remote control communication scenario for popular smart TVs}
  \label{fig:smartscreen_ele}
\end{figure}

While security researchers and TV manufacturers are making a concerted effort to strengthen smart TVs,
    we observed that they often ignore a new attack surface --- \textit{multi-channel remote control communication}.
Figure~\ref{fig:smartscreen_ele} depicts a typical application scenario:
    a smart TV simultaneously supports three types of remote controls using different signals, i.e., Consumer Infrared (IR)~\cite{infrared}, Bluetooth Low Energy (BLE)~\cite{ble}, and Wi-Fi.
In addition to 
remote controls provided by specialized
TV accessories,
    a smart phone can be used as a remote control when installing a \textit{companion app} developed by the TV manufacturer. By sending BLE and Wi-Fi signals, users can interact with the TV.
This \textit{companion app simulated remote control} is generally more powerful than those classical remote controls since it can fully make use of the resources of the host smart phone.


Although multi-channel remote control communication enhances easy-of-use and flexibility for smart TV users,
    it weakens security: a smart TV often treats its remote controls as benign accessories,
    and neither effectively authenticates their identities nor verifies data they send.
Unfortunately,
    most remote controls lack necessary protection,
    and thus attackers could easily impersonate a remote control or tamper the wireless traffic.
More seriously,
    to support enhanced features (e.g., playing video files from a 
    companion app simulated remote control),
    smart TV OSes add \textit{remote control interfaces} to handle sophisticated remote commands and execute privileged operations.
If the access control mechanisms of those interfaces are not well designed,
    attackers could simply abuse them to hijack the TV
    (i.e., monitoring the screen, displaying contents, and controlling the user interface (UI) of the TV).

\vspace{0.1cm}
\noindent
\textbf{\sysname Attack.}
In this paper, we present a new type of attack, \sysname,
    against multi-channel communication between a smart TV and its remote controls.
Unlike existing attacks that need to install a malicious app on the TV or exploit the TV OS,
    \sysname only reuses communications of remote controls to hijack the victim TV,
    making it more difficult to detect and prevent the attack.
We found that the root cause of this attack is a \textbf{multi-channel remote control mimicry} vulnerability
    (\sysname vulnerability for short).
In general,
    an \sysname vulnerability is a smart TV access control bug which allows an attacker to combine three types of wireless communications together
    (i.e., IR, BLE, and Wi-Fi)
    to circumvent the authentication and isolation policies of each single remote control. 
Then the attacker could abuse corresponding remote control interfaces to hijack the TV.
In fact,
    exploiting single remote control does not result in severe security threats.
However, by combining functionalities of multiple remote controls,
    one can design complex attacks.

To exploit an \sysname vulnerability,
    three consecutive steps are needed.
First,
    the attack utilizes less secure wireless channels (i.e., IR and BLE)
    to enforce a Wi-Fi provisioning~\cite{wifiprovision},
    a common procedure for smart TVs to receive credentials of a protected WLAN (i.e., SSID and password).
When inside the same WLAN, as
    most smart TVs would not check the remote control pairing requests,
    the attack leverages this weakness to actively bind a fake remote control to the TV.
Once the fake remote control is bound to the TV,
    the attacker then abuses the remote control interfaces to access TV resources and control the screen.


In comparison with attacks relying on meticulously crafted signals
    (e.g., leveraging inaudible voice commands to control the TV~\cite{zhang2017dolphinattack, roy2018inaudible, sugawara2020light}),
    the \sysname attack only uses common wireless technologies and is more general.
We conducted an empirical study against eight popular smart TVs from retail markets of the China, Japan, Korea and United States.
Our study showed all of them were vulnerable to the \sysname attack. Unlike attacks, such as BIAS~\cite{antonioli2020bias} against Bluetooth and KRACK~\cite{vanhoef2017key} against Wi-Fi,
    the \sysname attack does not aim to break any of the three wireless protocols used by remote controls.
Instead,
    it exploits the fact that during communications between the remote controls and the smart TV, because of usability considerations,  simplified security controls, or even no security controls at all, are applied.
We present a case study for the Samsung smart TV,
    which adopts the ISO/IEC 30118-1:2018 standard~\cite{ocf} to protect its remote control communication.
We show that a usability factor related to the Samsung \tool{SmartThings} companion app significantly reduced the crypto key randomness,
    and we constructed a practical brute-force attack to breach its DTLS-over-BLE and WebSocket-over-Wi-Fi communication between the TV and its companion app simulated remote control.


The main contributions of this paper are as follows:
\begin{itemize}

\item 
\textit{New Understandings.}
We systematically analyzed how the use of remote controls affects the security of popular smart TVs.
We show that design flaws of remote controls break the security assumptions of protection solutions currently deployed on wireless technologies such as BLE and Wi-Fi. 

\item 
\textit{New Attacks.}
We implemented the \sysname attack that affected 200 millions of popular smart TVs worldwide~\footnote{According to the shipments data reported by each smart TV manufacturer\cite{smarttvsta}.}.
Unlike attacks aiming at exploiting code vulnerabilities of TV OSes or apps, 
    \sysname attack only utilizes legitimate protocols and services.
Therefore, current protections are less effective against our attack. 
We also outline countermeasures for smart TV manufacturers and developers to mitigate the \sysname attack.
\end{itemize}

\section{Background}
In this section, 
we first give an analysis of smart TV characteristics by comparing them with other three types of devices.
Then, we describe common protection schemes of smart TVs, especially those to protect app simulated remote control communications.

\begin{table*}[!htp]
\renewcommand\arraystretch{1.5}
    \small
    \centering
    \caption{Comparison of implementation features among four types of consumer electronic devices}
    \label{tab:comparison}
\resizebox{2\columnwidth}{!}{
 \begin{tabular}{c|c|c|c|c}

 \toprule[1.5pt]

     & \textbf{Traditional TVs} & \textbf{IoT Devices} & \textbf{Smartphones} & \textbf{Smart TVs} \\
    \midrule[0.8pt]
    \textit{System} & - & Embedded OS/Bare & Android/iOS Metal & TV OSes~\cite{smarttvos} \\
    \textit{User Interface} & Menu & Web/Smartphone & Touchscreen & Desktop (Menu+TV app) \\
    \textit{Interaction} & Remote Control & Companion App & Screen-based Input & Remote Control + Companion App \\
    \textit{Communication} & IR & BLE/WiFi & BLE+WiFi & IR+BLE+WiFi \\

\bottomrule[1.5pt]
\end{tabular}}
\end{table*}
    
    
    
     


\subsection{Characteristics of Smart TVs}\label{sec:bg:char}
Smart TVs provide a variety of new features and functions for users, and thus their user experiences greatly differ from other devices, such as smartphones and laptops. 
A TV is considered ``smart'' when it has the following features:
    1) it relies on an OS to manage the hardware to process the displayed contents; 
    2) it can access online media resources through Internet connections;
    3) it supports multiple accessories that communicate through various transmission channels.
Compared with other widely used electronic devices including traditional TVs, smartphones as well as common IoT devices, smart TVs have the following major differences (see Table~\ref{tab:comparison} for a summary):

\vspace{0.2cm}
\noindent
\textbf{Systems}
Because of the limited resources (e.g. small memory, limited power), traditional TVs and IoT devices are usually not configured with a fully functional OS but just with a tailored embedded OS, or even a bare metal firmware with simple structures and functionalities. 
Smartphones, with more powerful hardware,
are equipped with OSes (Android or iOS) and support different types of apps.

Like for smartphones,  smart TV manufacturers often customize OSes (TV OSes) to adapt to the smart TV hardware and ``smart'' user interfaces.
Most smart TV manufacturers build their TV OSes on top of an existing OS, such as \tool{Android TV}~\cite{android_tv} developed by \tool{Google} or \tool{tvOS}~\cite{tvos} developed by \tool{Apple}.
In addition, smart TV apps, like smartphone apps, are provided to facilitate the use of smart TVs by  users.
Smart TV apps are mostly provided as pre-installed apps by smart TV manufacturers.
Some of TV OSes, however, also support the installation of third-party TV apps (often with proprietary TV app stores).

\vspace{0.1cm}
\noindent
\textbf{User Interface (UI)}
Generally, a traditional TV consists of a screen and a cable to display analog signals decoded media.
To support user interaction,
traditional TVs usually display menus on the screen for users to select.
User interactive inputs are limited to simple operations, such as switch-on/off and channel/volume tuning, and such operations are often conducted by the user with a remote control.
IoT devices, on the other hand, often lack a visible UI for operations.
Therefore, they usually rely on a remote web or smartphone based user interface to handle user inputs. For smartphones, the UIs are much more complex. With the help of touchscreen,
users can simply operate the smartphones with different multi-touch gestures. 

The UIs of smart TVs combine the features of UIs of traditional TVs and smartphones.
For usability consistency, most smart TVs still support a menu based operation style, 
but also support the use of TV apps (e.g., a media player app) to enhance the functionalities as well.

\vspace{0.1cm}
\noindent
\textbf{Interaction}
Regarding the input styles, the interactions between users and the four types of devices differ significantly. Traditionally, users operate the TV screen with a remote control, which only sends commands to the TV.
When using an IoT device, users usually rely on a companion app on a smartphone to send and receive messages. With respect to smartphones, the typical interaction approach is  touchscreen-based, while some smartphones also receive voice commands to fulfil certain functions.

Since users of smart TVs seldom touch the screen, most recent smart TVs still rely on the remote controls as their main accessories.
However, the remote control of a smart TV is ``smarter'' compared to that of a traditional TV.
It supports not only sending button-pressing commands but also sending voice commands via a short-range wireless communication channel.
In addition to the smart remote control, many manufacturers also provide companion apps that can be installed on smartphones, by which users can control the smart TVs from their smartphones. 
In particular, some companion apps not only allow the user to operate the TV remotely, but also allow the user to play the contents stored on the smartphone on the smart TV.

\vspace{0.1cm}
\noindent
\textbf{Wireless Communication}
A distinct feature of smart TVs is the use of multiple wireless communication channels.
As Figure~\ref{fig:smartscreen_ele} shows, the communication between a smart TV and its accessories (including remote controls and smartphones) utilizes three widely used wireless signals, 
i.e., \textit{Consumer Infrared}~\cite{infrared} (IR), \textit{Bluetooth Low Energy}~\cite{ble} (BLE), and \textit{Wi-Fi}~\cite{wifi}.
Commonly employed in traditional remote controls,
IR based short-range TV-Accessory communication is still supported by most smart TVs due to user experience considerations and compatibility issues.
Specifically, when a user presses a button on a remote control,  the remote control sends the corresponding IR signal to the smart TV.
The IR receiver on the TV then decodes the IR signal into instructions that the TV OS can understand.
Many smart remote controls (especially those with microphones to receive voice commands)
use BLE, which has a higher data transmission rate, to communicate.

Unlike remote controls, companion apps on smartphones tend to control the smart TV and access TV resources via Wi-Fi. Wi-Fi transmission not only has a high data rate but also adopts well-designed security specifications (e.g., WPA2 and WPA3~\cite{wpa}), and therefore is suitable for data transmission with strong protection requirements.
In comparison, IR and BLE lack strong authentication mechanisms.  IR communication does not need to authenticate the involved devices~\cite{zhou2019potential}, and BLE authentication suffered from pairing issues such as Man-in-the-Middle, Brute-force and Method Confusion attacks~\cite{ryan2013bluetooth, zegeye2015exploiting, melamed2018active, zuo2019automatic, zhang2020breaking, wu2020blesa, von2020method}.
As a result, IR and BLE based remote controls are restricted to fulfil a limited number of operations requiring privileges.

\subsection{Protections against Wireless Attacks}\label{sec:bg:protection}
Since complex hardware and software stacks are introduced into smart TVs,  a variety of vulnerabilities have been exploited against different components of the smart TV, such as the firmware and the browser~\cite{sidiropoulos2013smart, michele2014watch, bachy2015smart, bachy2019smart, mohajeri2019watching}.
In response, manufacturers and TV OS developers have built on techniques designed for smartphones protection and have adopted several well known defenses, such as Mandatory Access Control (MAC) and Address Space Layout Randomization (ASLR).
Nonetheless, a remarkable attack surface for smart TVs is their TV-Accessory wireless communication.
To protect smart TVs against remote wireless signals based attacks, the following measures are often employed.

\vspace{0.2cm}
\noindent
\textbf{Protection I: Network Isolation.}
When a user initially launches a smart TV, 
the user typically configures the TV to connect to a Wi-Fi network. At this stage, the smart TV often relies on the user to send the Wi-Fi credentials (i.e., the SSID and password of the WLAN) via its remote controls. Those credentials are often sent from remote controls to TVs via IR or Bluetooth, since at that time the Wi-Fi connection has not yet been established. After the network connection,
the smart TV is under the protection of WLAN isolation. Thus,  only authenticated devices are allowed to join the (WLAN) network and access TV resources.

\vspace{0.1cm}
\noindent
\textbf{Protection II: TV-Accessory Binding.}
In addition, the smart TV and its accessories (a remote control or a smartphone with a companion app) in the same WLAN need to complete a binding process before further remote interactions.
Conventionally, a binding process involves a mutual authentication between the smart TV and the accessory, which ensures the smart TV to be bound with the permitted accessories only.
Otherwise, attackers might be able to exploit the smart TVs by compromising other vulnerable smart devices (e.g., smart routers) in the same WLAN.

\vspace{0.1cm}
\noindent
\textbf{Protection III: Remote Interaction Validation.}
Finally, the remote user is not allowed to use resources of the smart TV arbitrarily.
Since a variety of remote user interactions supported by the smart TV require to access to sensitive resources
    (e.g., screen contents, system settings) or modify these resources,
    the smart TV  applies access control to all remote operations to check whether a request is allowed. 
Specifically, the TV OS introduces new interfaces to handle different remote operations sent by the user and perform permission checks. 
The permissions are granted to accessories after the binding phase, and when a resource interface is invoked by an accessory,
    the TV OS determines whether the accessory has been granted the specific permissions to access the resource. 
To distinguish different accessories, the smart TV generally distributes an access token to each accessory beforehand, and asks any remote request to attach the proper access token.

\begin{figure*}[!htbp]
\centering
\includegraphics[width=1\textwidth]{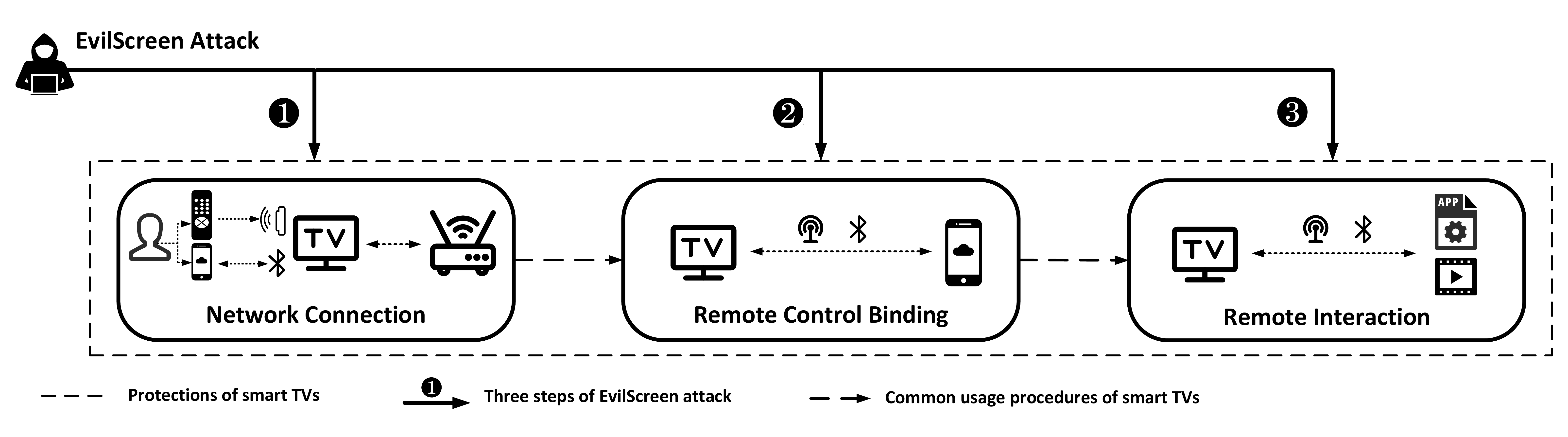}
\caption{A typical flow of \sysname attack: it conducts a consecutive three-step attacking process to circumvent common protection measures and finally hijacks the screen of the smart TV}

\label{fig:attackmodel}
\end{figure*}

\section{Threat Model}
Our attacker model is as follows.
\begin{itemize}
\item 
The smart TV OS is securely implemented. Thus the attacker cannot hijack and compromise the smart TV by exploiting the OS. In addition, we assume that no malicious apps were previously installed on the victim TV.

\item 
In order to analyze the smart TV and find potential security flaws, we assume that the attacker can purchase any products beforehand. Also we assume that, through his own analysis,
the attacker is able to gather necessary information, such as device configurations and companion app implementations, and extract the specific side channel information (e.g., certain range of mac address) to identify the model and brand of the victim TV~\cite{chen2019your}.

\item
The communication channels between the accessory and the smart TV are exposed to the attacker. 
The attacker can sniff network traffic and capture network packets.
However, we assume that the attacker is not able to circumvent existing communication channel protections.
For instance, data transmitted through the Wi-Fi channel is protected, that is, the attacker is unable to crack Wi-Fi credentials by launching brute-force attacks. 
\end{itemize}

Of course, in real-world scenarios, smart TVs of the same brand may be installed at various locations.
Depending on the location variants, the attacker model might also be slightly different.

\vspace{0.2cm}
\noindent
\textbf{Publicly Accessed Smart TVs.}
A large number of smart TVs are placed in public areas, such as shopping malls and gyms for commercial or entertainment purposes (e.g., advertising). Under such a scenario, the attacker can easily approach the publicly placed smart TV. Therefore, the attacker can not only monitor (and exploit) the remote communication between the TV and its accessories, but can also actively send malicious signals to the screen. However, if the attacker significantly changes the content on the screen, the attack is easily discovered. Thus attacks against publicly accessed smart TV must be stealthy.

\vspace{0.1cm}
\noindent
\textbf{Personal Smart TVs.}
As the personal smart TVs are generally placed in private spaces, such as homes or hotel rooms, the attacker cannot easily get close to them.
The attacker would then have to place a malicious device nearby (e.g., outside the door), or compromise a vulnerable device inside the private space, and use it as a relay device to launch the follow-up attacks.
We assume that the relay device can constantly sniff the victim network and actively send various types of signals (i.e., IR, BLE and Wi-Fi).
More importantly, most smart TVs do not actually power off themselves.
Instead, when the user sends a ``power off'' command to a smart TV, the smart TV only
turns off the screen and keeps its OS running.
Therefore, an attack can be launched during certain time periods when the TV is seldom used (e.g., late nights) without the user noticing anything. 
The attacker can then wait for the victim user to use the smart TV.


\section{The \sysname Attacks}
\sysname attack is a new type of wireless communication attack that exploits a type of \textbf{multi-channel remote control mimicry} vulnerabilities (\sysname vulnerabilities for short).
Abstractly,
    an \sysname vulnerability exploits three weaknesses in the architecture/implementation of smart TV systems to circumvent the three protections presented in Section~\ref{sec:bg:protection}.
The first weakness is that the victim smart TV supports various remote controls using different wireless signals,
    and there exists an \textit{implicit authentication dependency} between two or more remote controls.
The second weakness is that the UI of the smart TV OS fails to provide enough information for users to distinguish malicious remote controls from benign ones.
The third weakness is that the smart TV OS provides a series of interfaces for remote controls to access sensitive data on smart TV and execute high-privilege operations,
    while the access control of those interfaces is ill-designed.
By simultaneously leveraging these three weaknesses, a successful \sysname attack would allow the attacker to remotely monitor the screen and/or perform hijacking against the victim smart TV.

Unlike attacks that utilize low-level code implementation bugs to compromise smart TV systems (e.g., Android TV OS),
    the \sysname attack is a high-level access control circumvention attack.
A generalized \sysname attacking process consists of three consecutive steps:
    \ding{182} \textit{network isolation bypassing}, \ding{183} \textit{malicious remote control binding}, and \ding{184} \textit{remote interfaces abusing}, as shown in Figure~\ref{fig:attackmodel}.




\subsection{Network Isolation Bypassing}
To protect the smart TV from being connected by unauthorized devices, the smart TV is isolated by a secure WLAN. 
Although authentication schemes in smart TVs are securely designed and implemented, such authentication schemes have a major issue of \textit{authentication dependency}.
To be specific, 
    we found that Wi-Fi provisioning is commonly conducted by IR/BLE communications; however both IR and BLE communications are insecure.
Therefore, it is 
    vulnerable to passive and active attacks (e.g., man-in-the-middle attacks and impersonation attacks).

In order to bypass network isolation,
    an attacker could either retrieve authentication  credentials to connect a malicious device to the secure WLAN or force the smart TV to re-connect to a new malicious WLAN.
To launch the above attacks,
    we analyze IR/BLE communications to exploit the authentication dependency issues.

\vspace{0.2cm}
\noindent \textbf{Analysis.}
To identify authentication dependency issues,
    we capture IR/BLE wireless signals to explore the authentication schemes utilized by each type of signals. 
In advance, we check the user instruction and the technical manual of each smart TV to confirm which types of wireless channels are supported.
Then we execute WLAN provisioning. 
If either IR or BLE is supported,
    we examine whether they are used for distributing Wi-Fi credentials, and security of the distribution.

To analyze IR signals, we leverage an existing database~\cite{lirc} and mobile apps with IR remote control functions (e.g., Universal Remote Control~\cite{rc_app1}) to integrate the existing encoding and decoding methods.
Then an IR receiver is used to capture the IR signals from remote controls during network provisioning, and we decode the IR signals by utilizing the integrated decoding methods.
If Wi-Fi credentials (i.e., Wi-Fi SSID and password) are identified from IR signals, we consider the smart TV as suffering from authentication dependency issues.
We further verify if there is any authentication mechanism embedded in the IR channel.
In particular, 
    we build an IR simulator on top of an IR emitter~\cite{phoneswithir, orviboir} and use the simulator to send simulated IR signals in different encoding schemes. 
If the smart TV accepts these IR signals and executes the corresponding operations,
    we consider Wi-Fi credential distribution as \textit{authentication dependent} over a vulnerable IR channel that cannot defend against active impersonation attacks.

While analyzing BLE signals, we leverage \tool{TI CC1352 Development Board}~\cite{ticc1352} to sniff BLE packets. 
We first analyze the packets and examine the Secure Connection-bit in the BLE Pairing packets.
If the Secure Connection-bit is set to 0,
    it indicates that the devices are bound via \textit{Legacy Connection}, which is vulnerable to brute force attacks~\cite{ryan2013bluetooth}. 
When a \textit{Legacy Connection} vulnerable BLE scheme is identified, 
    we further utilize \tool{Crackle}~\cite{crackle} to decrypt the BLE packets and check whether there are any credentials included. 
Similar to IR channel analysis, 
    if BLE packets contain Wi-Fi credentials, 
    we consider credential distribution as \textit{authentication dependent} on an insecure BLE channel, which is vulnerable to passive man-in-the-middle (MITM) and brute force attacks.
Additionally, 
    we determine which BLE pairing mode is used by checking the MITM-bit and IOCaps~\cite{ble5core}.
If none of the MITM-bits in the both exchanged device pairing features is 1, the mode is \textit{Just Works}, in which no authentication mechanism is applied. 
In this case,
    we also regard Wi-Fi provisioning procedure as \textit{authentication dependent} on a BLE channel vulnerable to active impersonation attacks.

\vspace{0.2cm}
\noindent 
\textbf{Attacks.}
We launch either passive attacks or active attacks to compromise smart TVs.
When Wi-Fi credentials are identified from the vulnerable BLE communications with \textit{Legacy Connection}, 
    we use the extracted Wi-Fi credentials to connect a malicious device to the same WLAN. 
Alternatively,
    if the Wi-Fi provisioning procedure is \textit{authentication dependent} on a vulnerable IR or BLE channel without any authentication mechanisms against active impersonation attacks,
    we impersonate a legal user to compromise the smart TV. 
By connecting a malicious remote control with the TV, we are able to force the TV to reconnect a new malicious WLAN.

\subsection{Malicious Remote Control Binding}


Because of efficiency and usability reasons, 
manufacturers usually deploy light-weight, easy-to-understand, but insecure binding mechanisms. 
In order to bind a smart TV with its remote control, 
manual attestation is generally required. 
However, most smart TVs only display a device name and a binding token on the screen. 
This information is insufficient for users to distinguish whether the request is sent from a legitimate remote control or a malicious one; thus the binding mechanisms are vulnerable to impersonate attacks. 

Hence,  we investigate what binding information is displayed on the screen and exploit the transmission channels to identify whether the binding information is verified by the smart TV. 
    
\vspace{0.2cm}
\noindent
\textbf{Analysis.}
To exploit the protection schemes of remote control binding, 
    we execute UI differential analysis to investigate the binding information utilized by the smart TV and how the binding request is formed.  
First, 
    we use different smart remote controls with unique identifiers (e.g., series numbers and MAC addresses) to send binding requests to each smart TV manually.
Then we record the binding information displayed on the screen while using different remote controls. 
By comparing the information generated for different remote controls, 
    we examine whether the displayed information is invariant. 
The binding authentication is regarded as vulnerable if the display information is constant or only limited device information is provided. 



\vspace{0.2cm}
\noindent
\textbf{Attacks.}
After having obtained the binding information, 
    we modify the binding request to connect a malicious remote control with the smart TV. 
First, 
    we check whether the binding request is properly validated by the smart TV.
By monitoring network traffic, 
    we intercept communication packets to study the packet format.
If the binding request is transmitted over an insecure communication channel, 
    we can directly explore the packet format and further change the authentication-related fields containing binding information (e.g., ``username'', ``password'', ``device name'') with the legitimate remote control information and then send the modified packets to to the smart TV to request for binding.
For the binding requests protected by the SSL/TLS protocol,
    we use \tool{Burp Suite}~\cite{burp} to retrieve the binding request format and then replace the authentication-related fields with the legitimate information.
The forged request is then sent from a fake client.
If the smart TV accepts the request and displays an indistinguishable binding information on the screen,
    we consider such a binding mechanism as vulnerable.

Some smart TVs rely on binding tokens to protect against impersonate attacks. 
Nonetheless, 
    such a binding scheme is still vulnerable because the involved token is not well-protected. 
In particular, 
    a binding token may be broadcast to the remote control, or embedded in the companion app by default. 
If the token is broadcast, 
    we intercept the communication packets to obtain the corresponding token. 
Otherwise, 
    we reverse engineer the companion app to retrieve the token.

\subsection{Remote Interface Abusing}
As a smart TV stores a variety of sensitive resources (e.g., system setting, media files, user configuration),
the smart TV checks access permissions to avoid arbitrary resource access. 
Unlike personally owned devices (e.g., smartphones, labtops),  smart TVs commonly apply a coarse-grained access control to protect against unauthorized access because smart TVs are commonly shared by a group of people such as a family (private-use) or consumers (public-use).
Such an authorization scheme has \textit{permission check} weaknesses that can be exploited by an attacker to access resources without having the corresponding permissions. 

In order to check whether the permissions are properly granted, 
    we investigate the protocols for remote interactions and forge commands to access the unauthorized resources. 
Noted that we focus on analyzing Android companion apps because most users operate the TV smart features (e.g., screencast and screenshot) by using their smartphone.

\vspace{0.2cm}
\noindent 
\textbf{Analysis.}
To study the protocols for remote interactions, 
    we first capture network traffics transmitted between the smart TV and its companion app by using \tool{tcpdump}~\cite{tcpdump}.
By analyzing network packets,
    we identify the communication protocol (e.g., MQTT, HTTP and private application layer protocols over TCP or UDP) used for transmitting control commands by \tool{Wireshark}~\cite{wireshark}.
According to the communication protocol,
    we determine which standard APIs~\cite{java_api, linux_api} should be applied for sending and receiving data.
For example, 
    API \texttt{<java.net.Socket: java.io.OutputStream getOutputStream()>} is adopted for TCP communication, while \texttt{<java.net.DatagramSocket: void send(java.net.DatagramPacket)>} is used for UDP.

After retrieving network configurations, 
    we recover remote interaction protocols and identify the authentication fields. 
Specifically, 
    we utilize \tool{JEB}~\cite{jeb3} and \tool{IDA PRO}~\cite{idapro} to reverse engineer the companion app.
Starting from each argument of identified network APIs,
    we carry out a backward program slicing to identify the variables that are directly/indirectly data-dependent on the argument to determine how the argument is constructed. 
Given the Data Dependence Graph (DDG), we further identify the authentication variables that are related to access control. If the variable is assigned a constant or a value generated by a pseudo-random number generator or a timestamp, we consider the variable irrelevant to access control since these values do not contain any identity information.
Otherwise,  when a variable is assigned by a value related to the smart TV (e.g., binding credentials) or a return value of a memory-read function (e.g., \texttt{<android.content.Context: java.io.FileOutputStream openFileOutput\\
(java.lang.String, int)>}),
    we conclude that the variable is relevant to access control.

\vspace{0.2cm}
\noindent \textbf{Attacks.}
Given the variables that are related to access control, 
    we launch remote interface abusing attacks to execute sensitive operations such as screencast and screenshot. 
First, we extract how the operation commands are formed and then modify the values of these variables by using the dynamic instrumentation framework \tool{Frida}~\cite{frida}.
We further send the modified commands via a malicious device. 
If the smart TV is operated successfully without any warnings and we can access the unauthorized resources arbitrarily, 
    we regard the remote interaction access control of the smart TV as vulnerable.
For example, the attacker can send a screenshot command to the smart TV when a user is using the TV. Then the content displayed on the TV will be captured and leaked.

\section{Experimental Results}
We launched \sysname attacks to test real-world devices and reported observed security flaws.

\begin{table*}[!htp]
\renewcommand\arraystretch{1.5}
    \small
    \centering
    \caption{Implementation features of investigated smart TVs}
    \resizebox{2\columnwidth}{!}{
    \begin{tabular}{c|c|c|c|c|c}
    \toprule[1.5pt]
        \textbf{Manufacturer} & 
        \textbf{Model} & 
        \textbf{TV OS} & 
        \textbf{Remote Control (RC)} &
        \textbf{Wireless Signals} &
        \textbf{Companion App (CA)} \\ 

\midrule[0.8pt]
\tool{Samsung}         & \textit{UA55RUF60EJXXZ}        & \tool{Tizen OS}       & \textsc{BLE}       & \textit{IR + BLE + Wi-Fi}           & \tool{com.samsung.android.oneconnect}	\\
\tool{TCL}         & \textit{32L2F}        & \tool{TV$+$ OS}*       & \textsc{IR}       & \textit{IR + Wi-Fi}           & \tool{com.tnscreen.main}	\\
\tool{Hisense}       & \textit{32V1F-R}   & \tool{VIDAA OS}*  & \textsc{IR}       & \textit{IR + Wi-Fi}       &  \tool{com.hisense.ms.fly2tv}	\\
\tool{Xiaomi}         & \textit{Mi4A} & \tool{PatchWall}*     & \textsc{BLE}       & \textit{IR + BLE + Wi-Fi}   & \tool{com.xiaomi.mitv.phone.tvassistant} 	\\
\tool{Sony}        & \textit{KD-55X8566F}      & \tool{Android TV}    & \textsc{BLE}       & \textit{IR + BLE + Wi-Fi}          & \tool{com.sony.tvsideview.phonev}	\\
\tool{Skyworth}       & \textit{32X8}       & \tool{Coocaa OS}*      & \textsc{IR}       & \textit{IR + Wi-Fi}        & \tool{com.coocaa.tvpi} 	\\
\tool{LeTV}      & \textit{Q32}    & \tool{EUI}*      & \textsc{BLE}       & \textit{IR + BLE + Wi-Fi}           & \tool{com.letv.android.remotecontrol}	\\
\tool{KONKA}     & \textit{K32K6}          & \tool{YIUI}*  & \textsc{IR}       & \textit{IR + Wi-Fi}      & \tool{com.konka.MultiScreen}	\\


\bottomrule[1.5pt]
\end{tabular}}\label{tab:targetdeviceinfo}
\begin{tablenotes}
\item *: These smart TV manufacturers build their TV OSes on top of \tool{Android TV}~\cite{android_tv}.
\end{tablenotes}
\end{table*}

\subsection{Experiment Setup}
To investigate whether protections are securely implemented in real-world smart TVs, we tested eight smart TVs manufactured by different well-known manufacturers~\cite{tv_shipment}, i.e., \tool{Samsung}, \tool{TCL}, \tool{Hisense}, \tool{Xiaomi}, \tool{Sony}, \tool{Skyworth}, \tool{LeTV}, and \tool{Konka} from China, Japan, Korea and United States.
Shipments of smart TVs from those manufacturers range from 
    7 million (\tool{Konka}) to 48 million (\tool{Samsung}).

All these eight smart TVs are equipped with smart TV OSes and remote controls.
Details about each smart TV are listed in Table~\ref{tab:targetdeviceinfo}.
Most manufacturers customize their smart TV OSes on top of \tool{Android TV}~\cite{android_tv}.
We list technical details of each smart TV and its remote control in Table~\ref{tab:attackresults}.
By default, all eight smart TVs support IR communications and four of them (i.e., \tool{Samsung}, \tool{Xiaomi}, \tool{LeTV}, \tool{Sony}) also support BLE communications. 
In our analysis of the remote controls, we found that not all of them are based on IR. 
Only \tool{TCL}, \tool{Hisense}, \tool{Skyworth} and \tool{Konka} smart TVs provide IR-based remote controls, whereas the others only provide BLE-based remote controls.
An interesting observation is that although the official manuals of \tool{Samsung} and \tool{Xiaomi} TVs did not mention about receiving IR signals, 
we could operate their smart TVs by sending IR commands. This indicates that IR receivers are still integrated in these two smart TVs. 

We further analyzed each smart TV by launching the \sysname attack.
We also reported all the discovered flaws and the consequences to the corresponding manufacturers, and \tool{Xiaomi} responded with a confirmation before the submission.

To comply with research ethics guidelines,  all the experiments were performed on our own devices and conducted in our lab testing environment.

\subsection{Network Isolation Bypassing}
\label{sub:eval:auth}
We exploited Wi-Fi provisioning of smart TVs and successfully compromised all the smart TVs, 
    that is, all the smart TVs are authentication dependent on vulnerable channels when provisioning Wi-Fi.
Hence, the attacker can utilize other vulnerable wireless channels to crack network isolation.
The detailed experimental results are described in what follows.

All eight smart TVs support the IR based Wi-Fi provisioning. 
By sending the tampered IR signals to the smart TV, we successfully forced all these smart TVs to reconnect to a new (malicious) WLAN.
These results show that WLANs can be bypassed because all these IR channels are vulnerable to impersonation attacks.
With respect to the BLE channels, the smart TVs of \tool{Samsung}, \tool{Xiaomi}, \tool{LeTV}, and \tool{Sony} can provision Wi-Fi through BLE remote controls. However, all the involved Wi-Fi credentials were distributed insecurely. We discovered that these four smart TVs adopt the \textit{Legacy Connection} scheme to bind with the remote controls; hence their BLE channels are vulnerable to brute force attacks. 
We then ran \tool{Crackle} to decrypt all the communication packets and successfully recovered all the transmitted data including Wi-Fi authentication credentials. 
We further found that all the four smart TVs are implemented with \textit{Just Works} pairing mode for BLE remote controls binding; as a result, they are also vulnerable to active impersonation attacks.
Thus,
    we successfully connected to these four smart TVs and forced them to reconnect a new malicious network from a fake BLE client.

We found an exceptional case in using BLE for Wi-Fi provisioning: 
\tool{Samsung} introduces a companion app, \tool{SmartThings}, by which users can provision Wi-Fi. 
The app still sent encoded commands via the BLE channel,
    and \tool{Samsung} specifically designed a solution with a customized DTLS protocol to protect it. 
Nonetheless, we still found a security flaw of this BLE protection, which we discuss in Section~\ref{sec:eval:attack:case}.

\subsection{
Malicious Remote Control Binding
} \label{sec:eval:binding}
We conducted a UI differential analysis to analyze the binding information displayed on the smart screen. 
The results demonstrated that none of these smart TVs implemented a correct binding mechanism because only limited device information was provided and some of the binding information was not even protected. 
Referring to the binding information, 
    we successfully cheated all users and bound malicious remote controls with the smart TV.

\subsubsection{Remote Control Binding}
All the eight smart TVs supported either IR-based or BLE-based remote controls. 
Unfortunately, neither of them was secure (refer to Section~\ref{sub:eval:auth}).
They all silently connected with the TV without prompting any connection information or warning.

\subsubsection{Companion App Binding}
We analyzed the binding schemes between companion apps and smart TVs to explore the displayed binding information and whether the binding scheme is protected by secure identity validation.

\vspace{0.2cm}
\noindent
\textbf{Information Display.}
Only \tool{Sony} and \tool{Samsung} TVs displayed binding information on their screens. 
Specifically, \tool{Sony} TV showed the device name,  a pseudo-random pincode, and the remaining time. 
Instead \tool{Samsung} TV displayed a pseudo-random pincode (in DTLS over BLE communication) or directly prompted an alert with the device name for users to decide whether to confirm the device binding (in WebSocket over Wi-Fi communication). 
However, such displayed information was insufficient for users to pinpoint each unique device. 
Even worse, the other smart TVs (i.e., \tool{TCL}, \tool{Hisense}, \tool{Xiaomi}, \tool{Skyworth}, \tool{LeTV} and \tool{Konka}) accepted the binding requests without showing any alert on their screens, and thus legitimate users were unaware of malicious connections.

\vspace{0.1cm}
\noindent
\textbf{Connection Authentication.}
We successfully sent a forged binding request to connect a malicious remote control to the smart TVs of \tool{TCL}, \tool{Xiaomi}, \tool{Skyworth}, \tool{LeTV}, and \tool{Konka}.
By inspecting the communication protocols, we found that these smart TVs customized their own protocols without verifying identities of remote controls.  
For \tool{Hisense} TV, its binding credential (i.e., username and password) was embedded in the companion app; thus the attacker could obtain the credential by reverse engineering the app and connect with the TV.
    
\tool{Sony} and \tool{Samsung} supported two types of connection authentication.
In particular, \tool{Sony} supported BLE-based binding credential distribution. To bind the remote control with the smart TV, \tool{Sony} TV generated a token (i.e., a pseudo-random number) for each binding request and then broadcast it via Bluetooth. 
When its companion app received the token, it automatically connected with the smart TV without notifying the user.
Unfortunately, we inspected these tokens and found that were transmitted in plaintext; thus, any apps with the BLE permission could also receive the password and complete the binding.
If Bluetooth was turned off, \tool{Sony} displayed a pseudo-random pincode in four-digit and waited for the user to type in the pincode from the app. 
However, the token was transmitted through HTTP in plaintext, which is vulnerable to MITM attacks.     
Even though a secure network connection was established, attackers could still crack the four-digit token by launching brute-force attacks.
Similarly, \tool{Samsung} TV also displayed a pseudo-random pincode in eight-digit when Bluetooth of the companion app was turned on; however the pincode needed to be filled in manually. 
On the other hand, the companion app could send its binding request through Websocket over Wi-Fi communications. 
Nevertheless, \tool{Saumsung} TV created the TLS connection without checking its certificate,
which is vulnerable to impersonation attacks.

\begin{table*}[!htp]
\renewcommand\arraystretch{1.5}
    \small
    \centering
    \caption{Implementation features of remote controls related to our analyzed smart TVs}
 \label{tab:attackresults}
 \resizebox{2\columnwidth}{!}{
 \begin{tabular}{c|c|c|c|c|c|c|c|c}

 \toprule[1.5pt]

\multirow{2}{*}{\textbf{TV}} & 
\multirow{2}{*}{\textbf{\begin{tabular}[c]{@{}c@{}}Wi-Fi\\Provisioning\end{tabular}}} & 
\multicolumn{3}{c|}{\textbf{Remote Control Binding}} & 
\multicolumn{4}{c}{\textbf{User Interactions}} \\ \cline{3-9}
 & & \textbf{Channel} & \textbf{Protocol} & \textbf{Attestation}
 & \textbf{CA as RC} & 
 \textbf{\begin{tabular}[c]{@{}c@{}}TV App Operation\end{tabular}} &
 \textbf{Screencast} & \multicolumn{1}{c}{\textbf{Screenshot}} \\

 \midrule[0.8pt]

\multirow{2}{*}{\tool{Samsung}} & \multirow{2}{*}{IR + BLE + CA} & BLE &  UDP(DTLS) & Pincode & \multirow{2}{*}{\cmark} & \multirow{2}{*}{\xmark} & \multirow{2}{*}{\xmark} &  \multirow{2}{*}{\xmark} \\ \cline{3-5}
 & & TCP & WebSockets & Prompt+Confirmation & & & & \\ \hline

\tool{TCL}      & IR        & TCP   & Proprietary   & User operating    & \cmark & Open/Install/Uninstall & \cmark & \cmark \\ \hline
\tool{Hisense}  & IR        & TCP   & MQTT          & Password          & \cmark & Open/Install/Uninstall & \cmark & \cmark \\ \hline
\tool{Xiaomi}   & IR + BLE  & TCP   & Proprietary   & User operating    & \cmark & Open/Install/Uninstall & \cmark & \cmark \\ \hline
\tool{Sony}     & IR + BLE  & TCP   & Proprietary   & Pincode/password  & \cmark & Open & \xmark & \xmark \\ \hline
\tool{Skyworth} & IR        & TCP   & WebSocket     & User operating    & \cmark & Open/Install/Uninstall & \cmark & \cmark \\ \hline
\tool{LeTV}     & IR + BLE  & UDP   & Proprietary   & User operating    & \cmark & Open & \xmark & \cmark \\ \hline
\tool{Konka}    & IR        & TCP   & Proprietary   & User operating    & \cmark & Open/Install/Uninstall & \cmark & \cmark \\ 

\bottomrule[1.5pt]

\end{tabular}
}
\end{table*}



\subsection{
Remote Interface Abusing
}\label{sec:analysis:hijacking}

By analyzing remote user interfaces supported by each smart TV, we compromised all the smart TVs successfully and accessed unauthorized resources arbitrarily. Our results thus indicate that all smart TVs did not grant permissions properly.

\subsubsection{Screen Operation Through Companion app}
The companion apps of all tested smart TVs could be used as a remote control to for controlling the cursor.
Except for \tool{Samsung} TV, the other smart TVs could also perform app operations (i.e., execution, installation, and uninstallation).
By analyzing the communication traffic, we noticed that commands sent by the companion apps of \tool{Hisense}, \tool{Xiaomi}, \tool{Sony}, \tool{LeTV} and \tool{Konka} were transmitted in plaintext.
Although the \tool{TCL} companion app used the AES algorithm to encrypt the transmitted commands, we found that its encryption key was encoded in the companion app; thus attackers could retrieve the key by reverse engineering. 

Moreover, we discovered that the smart TVs of \tool{TCL}, \tool{Hisense}, \tool{Xiaomi}, \tool{Skyworh}, \tool{LeTV} and \tool{Konka} accepted all commands without checking their validity. 
Therefore, we tampered the commands to execute malicious operations such as malware installation. 
Although \tool{Sony} TV verified the commands by checking a cookie shared beforehand, attackers could extract the cookie by intercepting communication packets. By inspecting the packets manually, we found that the commands of app downloading sent by the companion apps of the \tool{Hisense} and \tool{Konka} TVs contained URLs. 
The smart TVs further downloaded app installation packets from the URLs instead of official app stores. Thus, we replaced the URL with the one of a malicious app installation packets and sent it to the smart TV. The smart TV successfully downloaded and installed the malicious app.

\subsubsection{Screencast}
Five smart TVs, i.e., \tool{TCL}, \tool{Hisense}, \tool{Xiaomi}, \tool{Skyworth} and \tool{Konka}, provided media files (i.e., video, photos, and music) and documents (e.g., ppt, pdf, txt) casting service (from the smartphone to the TV).
Instead of transmitting files directly,
    URLs for downloading the files were delivered to the smart TVs.
All these smart TVs displayed the received files directly without checking whether the sender is authorized for screencast. 
Apart from the \tool{TCL} TV, the screencast procedure of the other smart TVs is not protected, that is, the URLs were not encrypted during transmission and the integrity of these URLs was not verified. 
For the \tool{TCL} TV, the communications were encrypted by the AES algorithm whose encryption key is embedded in the companion app. 

Not only did the smart TVs suffer from remote interaction abuse, so did the corresponding companion apps.
Therefore, we could easily access the sensitive files on the smartphone by capturing the packets for the screencast service to obtain the URLs.

\subsubsection{Screenshot}
Six smart TVs (i.e., \tool{TCL}, \tool{Hisense}, \tool{Xiaomi} \tool{Skyworth}, \tool{LeTV}, and \tool{Konka}) support interfaces for screenshot. 
Similar to screencast, the
\tool{TCL}, \tool{Xiaomi}, \tool{Skyworth}, and \tool{Konka} TVs sent a URL from which to download the screenshot instead of sending the screenshot images. 
The corresponding companion apps further obtained the screenshot images through the received URLs.
On the contrary, \tool{Hisense} and \tool{LeTV} TVs directly transmitted the screenshot images back to their companion apps. 
In addition, \tool{Xiaomi} TV provided an interface which synchronizes the screen content to the companion app.

Nonetheless, we successfully launched screen hijacking attacks by continuously sending screenshot requests and then monitoring the screen contents watched by the user. Our results show that none of them provides authorization mechanism to protect screenshot images.

\subsection{Case Study: Samsung Smart TV}\label{sec:eval:attack:case}
Considering \tool{Samsung} smart TV as an example, we now describe in details how an \sysname attack was launched to exploit security flaws in the smart TV remote control binding.
In particular, the binding between the companion app, \tool{SmartThings}\footnote{\tool{SmartThings} is an official companion app developed by \tool{Samsung} for home automation. It can connect with \tool{Samsung} smart devices and control them. These devices include bulbs, speakers and smart TVs, etc.}~\cite{smartthings}, and the TV is regulated by the Open Connectivity Foundation (OCF) security specification of ISO/IEC 30118-1:2018~\cite{ocf}; however the security controls implemented based on this specification are not adequate due to the presence of ``smart'' user interfaces.
The \sysname attack was able to crack both device binding mechanisms, that is, BLE based binding and Wi-Fi based binding.

\begin{figure}[!htp]
\centering
\includegraphics[width=0.4\textwidth]{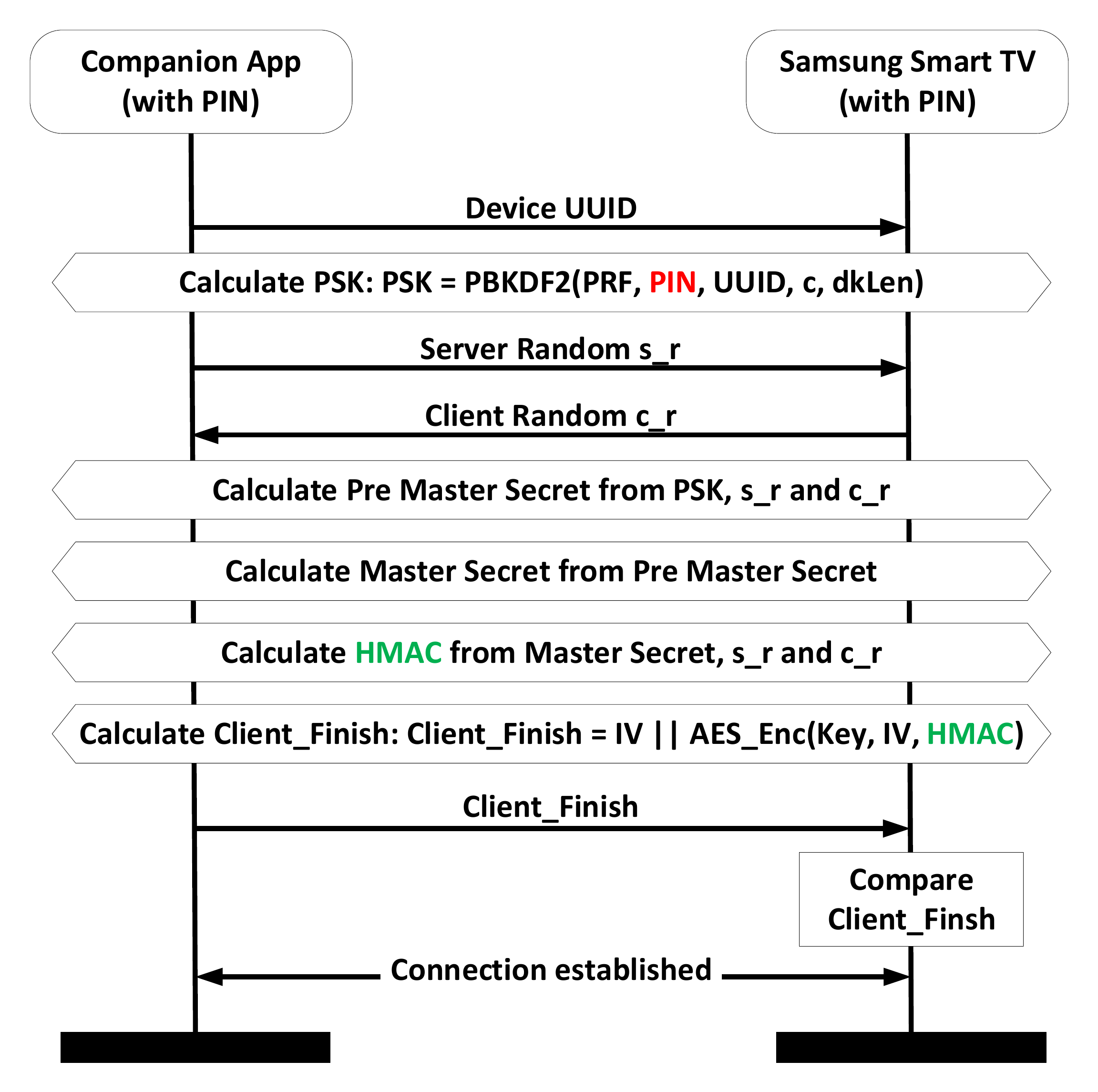}
\caption{Connection establishment between Samsung Smart TV and its companion app}
\label{fig:ssg_bind}
\end{figure}

\subsubsection{BLE based Binding Attack}
The communication protocol between 
\tool{SmartThings} and the smart TV is a customized DTLS protocol, built on top of Bluetooth 4.2.
The detailed binding authentication is illustrated in Figure~\ref{fig:ssg_bind}. 
First, \tool{SmartThings} sends its device UUID as a binding request to the smart TV.
Such a request is sent over an unprotected BLE channel. When the request is received, the smart TV generates a pseudo-random number (PRN) of eight digits and displays the number on its screen. 
The smart TV then regards the eight digits PRN as a pincode and calculates a pre-shared key (PSK) according to the following expression:

\begin{equation}
\small
     PSK = PBKDF2(HMAC\_SHA256, PIN, UUID, c, dKLen)
\end{equation}

\noindent
The calculation relies on the PBKDF2 algorithm. 
Within this algorithm,  the only secret field is the pincode (i.e., the PRN).
Simultaneously, the user types in the same pincode on \tool{SmartThings} to generate the same PSK. 
When both PSKs are confirmed to be the same,  the smart TV and the user generate two pseudo-random numbers, \textit{s\_random} and \textit{c\_random}, respectively, and further calculate a pre-master secret (PMS) and a master secret (MS) according to the following expression:
    
\begin{equation}
\small
\begin{split}
    PMS &= TLS\_ECDHE\_PSK(PSK, s\_random, c\_random) \\
    MS &= PRF(HMAC\_SHA256, PMS, Padding)
\end{split}
\end{equation}

\noindent
for generating a standard TLS session key. 
Finally, \tool{SmartThings} sends a message containing the HMAC-SHA256 digest to establish the TLS connection for binding credential distribution. 

\begin{figure}[tp]
\centering
\includegraphics[width=0.45\textwidth]{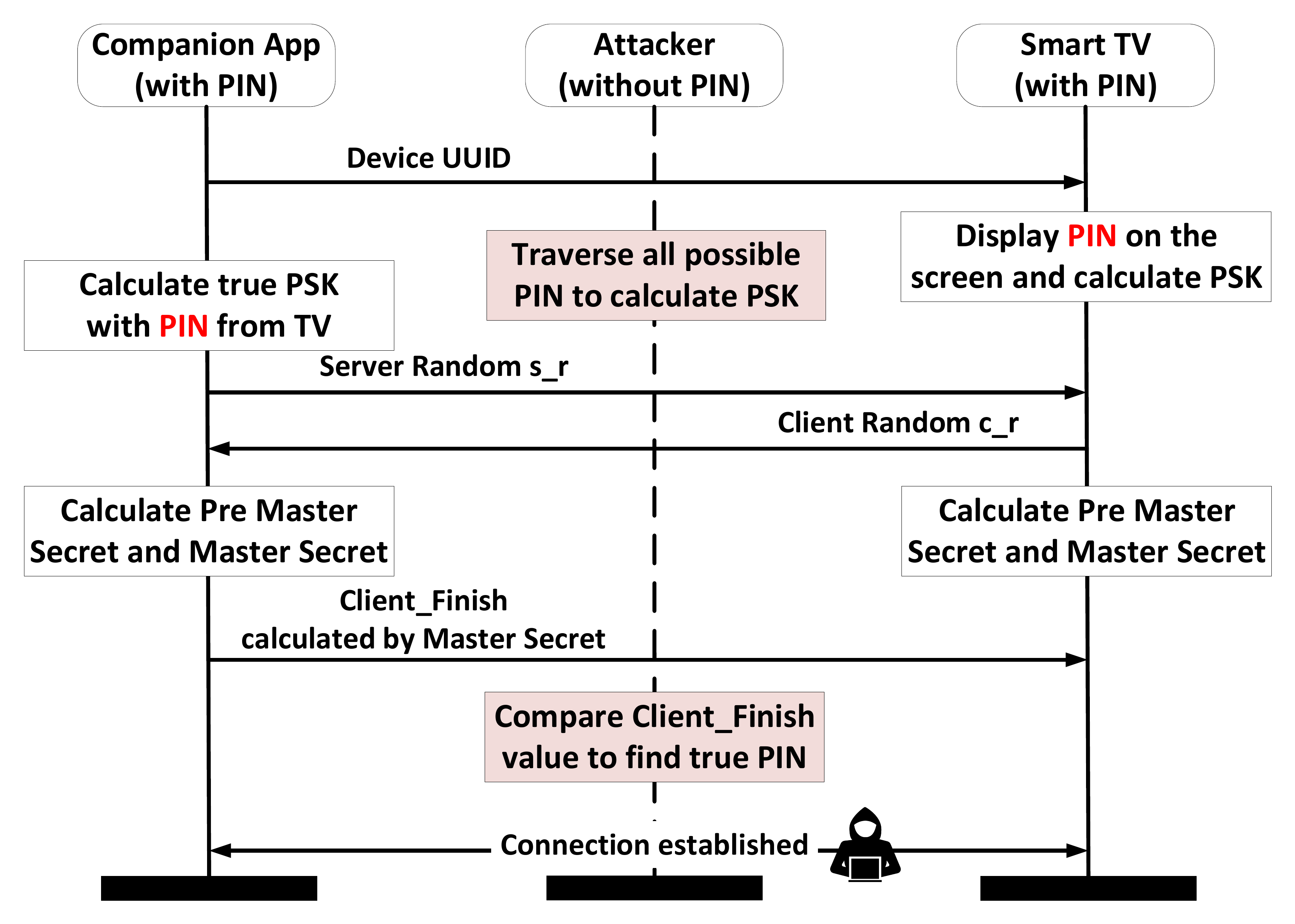}
\caption{MITM Attacks on Samsung Smart TV and its companion app}
\label{fig:ssg_attack}
\end{figure}

The only issue in this authentication protocol is the usage of the eight digits pincode. 
The security specification in OCF is to choose a PRN with enough entropy as the input of the PBKDF2 algorithm. However,  \tool{Samsung} smart TV reduces the search space against PSK to $10^8$ which is breakable within a short period of time. 
To exploit such a flaw, we launched a MITM attack again the binding authentication demonstrated in Figure~\ref{fig:ssg_attack}. We first monitored the BLE communication between \tool{SmartThings} and the smart TV. When a device UUID was detected, 
we pre-calculated all possible PSKs within the search range of $10^8$ and the corresponding MSs. 
By using the Client\_Finish packet, we tried all the possible PSKs as the secret key to decrypt the packet until an effective key is identified.

We conducted the key search by utilizing a 2080 Ti GPU to speed up the key search with \tool{hashcat}~\cite{hashcat}.
The entire guessing took only 40 seconds on average, which shows that  the attacker definitely has enough time to retrieve the PSK and reconnect the smart TV to a malicious device. Although the PSK cannot be cracked within a restricted period, we could still obtain the PSK by launching offline attacks and further decrypting the transmitted messages. 

\subsubsection{Wi-Fi based Binding Attack}
Prior to binding, \tool{SmartThings} and the smart TV were connected to the same WLAN. 
\tool{SmartThings} first sent a token request to the smart TV through a Websocket communication using the TLS protocol.
The request with a specific binding URL was constructed according to the format \texttt{wss:\/\/\{IP\}:\{PORT\}?name=\{DEVICE\}\&token=\{TOKEN\}}.
When the URL was first accessed, 
    a window with the device information popped up on the TV screen and asked whether the mobile device was allowed to connect to the TV.
If it was allowed, a token was then sent back to \tool{SmartThings}.

Nevertheless, the device name displayed on the TV screen can be obtained by the attacker easily. Then we utilized a malicious device to request for the TLS connection. To construct a WebSocket request, we modified the \textit{name} field by using the displayed device name. As there is no warning message indicating to the user that a remote control has connected, the user
will be easily misled (by thinking the previous connection to be unsuccessful) to accept the request and connect the smart TV with the malicious device.

\subsection{Discussion}
Although several security flaws of the smart TVs were exploited, some constraints may limit the capabilities of the \sysname attack.

\vspace{0.2cm}
\noindent
\textbf{Distance constraint.}
The \sysname attack targets the wireless communication of IR, BLE and Wi-Fi. 
To exploit IR and BLE communications, the attacker is required to approach the smart TV within a specific distance; thus he/she needs to come up with multiple ways to get close to the victim's smart TV. 

\vspace{0.1cm}
\noindent
\textbf{Prerequisite information.}
Since the \sysname attack needs to be launched stealthily, the attacker needs to learn detailed information of a smart TV, e.g., device name, BLE pairing mode, displayed binding information by either purchasing the same brand of smart TVs or (physically/virtually) entering the private property to check the corresponding smart TV.

\vspace{0.1cm}
\noindent
\textbf{Activity constraint.}
Attacks' capability will be affected if the smart TV is switched off, in which case the smart TV is disconnected from any wireless channels.
In this case, the attack is not possible as the attack requires at least one the wireless channels on the smart TV to be active.
However,  when the  smart  TV  is  turned  on, by using \sysname  the attacker can not only eavesdrop user’s private data but also actively attack the victim TV.

\section{Countermeasures}

Based on our findings, we suggest that TV manufacturers to improve their protection schemes as follows.

\vspace{0.2cm}
\noindent
\textbf{Restrict sensitive operations via vulnerable channels.}
The major issue of the smart TV is caused by the involvement of multiple wireless channels.
Among the supported wireless channels, IR and BLE are the most vulnerable ones; thus it is essential to avoid transmitting sensitive information (e.g., authentication credentials) through these vulnerable channels.

\vspace{0.1cm}
\noindent    
\textbf{Limit the number of connected remote controls.}
It is essential for manufacturers to limit the number of remote controls that are allowed to connect with the smart TV. 
For instance, when two remote controls are originally provided by a manufacturer, the smart TV can then be limited to bind with remote controls up to two.
The best would be to bind with one remote control only each time. 

\vspace{0.1cm}
\noindent
\textbf{Design multi-factor authentication.}
Instead of relying on the binding credential only, the smart TV can authenticate the remote control through multiple factors such as both binding credential and user confirmation. 
For instance, the smart TV could distribute a binding credential to the remote control and provide a reminder for the user with the binding details simultaneously. 
Furthermore, it is essential to strengthen the credential protection scheme such as utilizing mutual authentication and end-to-end encryption.

\vspace{0.1cm}
\noindent
\textbf{User-involved authorization.}
A fine-grained authorization scheme (similar to smartphones) is necessary to protect the remote control UIs. 
While designing the smart TV OSes, the manufacturer needs to constraint the operations that can be executed by the remote controls. 
Moreover, before launching sensitive operations such as screencast and screenshot, certain sensitive permissions need to be required again.

\section{Related Work}

\taitou{Smart TV Security}
Previous research has discovered many security issues in smart TVs.
Niemietz \textit{et al.} \cite{niemietz2015not} analyzed the authentication procedures of smart TV apps and observed that users' accounts can be hijacked via eavesdropping, physical access, and installation of malicious TV apps.
Moghaddam \textit{et al.} \cite{mohajeri2019watching} focused on the privacy practices of Over-the-Top streaming devices (e.g., smart TVs) and discovered that these devices collect users' private information and behavior habits for advertising and tracking. 
In order to address the problem of sensitive user information leakage, Kang \textit{et al.} \cite{kang2017obtain} proposed the first EAL2 certification worldwide to improve the security and reliability of Smart TVs. 
Such previous work, however, mainly focuses on analyzing the connection security without considering the procedure of establishing the connection. We not only consider the entire operation procedure including authentication, binding, and operation, but also discuss the severity of each security flaws.

Other research exposed attack surfaces and exploitable vulnerabilities of smart TVs.
David \cite{voicecommand} discussed the security of smart TV voice commands.
Mich{\'e}le \textit{et al.} \cite{michele2014watch} explored a weakness in the multimedia layer of TV which allow attackers to gain full control without a physical access to the TV.
Though our work also discovered security issues of smart TVs, we revealed the attack surfaces by focusing on the security of remote interaction.

Apart from the connection issues, Bachy \textit{et al.} \cite{bachy2015smart, bachy2019smart} contributed to the security of heterogeneous networks adopted by smart TVs.
They considered the security threats of Internet communications of Digital Video Broadcasting,
    the Asymmetric Digical Subscriber Line (ADSL) and further identified weaknesses that may lead to firmware modification and traffic fraud.
Differently, we focus more on the security of user interactions through multiple wireless networks
    (i.e., IR, BLE and WLAN) that are used frequently for TV users.
our attack aims to hijack the TV, which not only allow the attacker to eavesdrop the traffic, but also to monitor the screen and change the system UI.

\vspace{0.2cm}
\taitou{Short-Range Wireless Communication Security}
Short-range wireless techniques (e.g., IR and BLE) are widely used for smart devices communication. 
However, these techniques are implemented insecurely. 
Zhou \textit{et al.} \cite{zhou2019potential} investigated the potential security issues in IoT devices supporting infrared remote control and observed sensitive data leakage.
Ryan \textit{et al.} \cite{ryan2013bluetooth} proved that some security flaws in BLE could make it easy for attackers to implement eavesdropping attacks.
In addition, Garbelini \textit{et al.} \cite{garbelini2020sweyntooth} developed a systematic automated fuzzing framework for BLE protocol to discover insecure implementation behaviours.

Besides, prior studies also focused on the security of Wi-Fi schemes, e.g., Wi-Fi provisioning. 
Li \textit{et al.} \cite{li2018passwords} conducted a security analysis against eight different Wi-Fi provisioning solutions and indicated that unsafe transmission in Wi-Fi smart configuration could lead to password disclosure and other issues. 
Liu \textit{et al.} \cite{liu2019tell} proposed a new Wi-Fi connection method based on audio waves.
Wang \cite{kang2019u2fi} also provided a solution for IoT provisioning scheme with universal cryptographic tokens.
Though such previous work has raised security issues for short-range communication also present in smart TVs, 
    exploiting these wireless channel vulnerabilities alone may not cause great security threats.
Instead, an \sysname attack, by combining vulnerable wireless communications with various remote interfaces provided by smart TVs, 
    could hijack a victim smart TV and monitor user behaviors and habits.

\vspace{0.2cm}
\taitou{IoT Devices Security}
Other research~\cite{shoshitaishvili2015firmalice, costin2014usenix, david2018firmup} focused on IoT device security by analyzing firmware. 
Such research has identified different types of vulnerabilities of firmware images though symbolic execution, arbitrated emulation and semantic procedure similarity comparison respectively.
Some other prior work~\cite{wang2019looking, chen2018iotfuzzer, redini_diane_21} focused on IoT device security via the companion apps. 
They proposed that the security of IoT devices is reflected in their mobile companion apps to a certain extent; thus, companion app analysis could assist n device analysis, such as device components similarity and automatic fuzzing.
Moreover, due to the diversity of application scenarios and functional requirements, specific types of IoT devices may suffer from their own security issues. Hence, some work~\cite{ho2016smart, muller2017sok, alrawi2019sok} has analyzed the security of specific IoT devices, including smart locks, printers and home-based devices, etc.
In addition, some researchers also considered the security of communication between devices, clouds and apps, such as remote binding~\cite{chen2019your}, device pairing~\cite{han2018you, sethi2019misbinding}, messaging protocols~\cite{jia2020burglars} and the interactions~\cite{zhou2019discovering}.

The above work puts emphasis on the security of resource-restricted IoT devices that are unable to provide secure protections and thus need to rely on the cloud. 
Unlike IoT devices, 
    a smart TV is usually configured with enough resources to support various functions.
The main scope of our work is on the security of local communication between smart TVs and remote controls according to the smart TV special features and usage scenarios.
Moreover,
    we utilized several types of remote control interfaces to circumvent smart TV security protections, rather than exploiting vulnerabilities of one single channel.

\section{Conclusion}
In this paper, we systematically analyzed the security of wireless communications between smart TVs and their remote controls,
    and based on this analysis proposed a new attack,
    \sysname attack,
    which exploits insecure multi-channel remote control communication.
By executing different remote control commands with multiple wireless channels in a sophisticated way,
    our proposed \sysname attack allows the attacker to access and modify resources on the victim TVs.
We have reported security issues of eight popular smart TVs, which are vulnerable to \sysname attack, to the corresponding manufacturers,
    and suggested countermeasures to help address these issues.



\ifCLASSOPTIONcaptionsoff
  \newpage
\fi

\bibliographystyle{IEEEtran}
\normalem
\bibliography{reference}

\end{document}